\documentclass[italian,english]{article}
\usepackage[latin9]{inputenc}
\usepackage{color}
\usepackage{esint}

\newcommand{\lyxaddress}[1]{
\par {\raggedright #1
\vspace{1.4em}
\noindent\par}
}

\usepackage{babel}

\begin{document}

\title{\textbf{On the correctness of Relative Time Dilatation in Special
Relativity in vacuum: a rebuttal against the claims by Arthur Boltcho
in European Journal of Scientific Research ISSN 1450-216X Vol. 44
No. 4 (2010), pp. 610-611}}

\author{\textbf{Christian Corda }}

\maketitle

\lyxaddress{\begin{center}
International Institute for Theoretical Physics and Mathematics Einstein-Galilei,
via Bruno Buozzi 47, Prato ITALY and Institute for Basic Research,
P. O. Box 1577, Palm Harbor, FL 34682, USA %
\footnote{\begin{quotation}
\emph{Partially supported by a Research Grant of The R. M. Santilli
Foundations Number RMS-TH-5735A2310}
\end{quotation}
}
\par\end{center}}

\begin{center}
\textit{E-mail addresses:} \textcolor{blue}{cordac.galilei@gmail.com}
\par\end{center}
\begin{abstract}
In his recent paper published in the European Journal of Scientifi{}c
Re­search 44, 4, 610-611 (2010), the author, Arthur Boltcho, claims
to have found a mathematical disproof of relative time di­latation
of Special Relativity Theory (SRT). In this letter we show that the
supposed mathematical disproof of rel­ative time dilatation of SRT
is totally wrong and that Arthur Boltcho demonstrated nothing. The
errors by Boltcho arise from a strong misun­derstanding and confusing
the concept of \textquotedblleft{}moments\textquotedblright{} and
time intervals in the framework of SRT. 
\end{abstract}
The author of ref. \cite{key-1} claims to release a mathematical
disproof of relative time dilatation of SRT. The situation in \cite{key-1}
is as it follows. An inertial stopwatch A is located at one of the
points of a circular trajectory along which a stopwatch B is moving
with constant velocity below that of light. These stopwatches are
activated simultaneously when they meet and, after one turn, at their
next meeting, they are deactivated simultaneously (valid certainly
for any observer). Then, the author of \cite{key-1} claims that,
verbatim:

{}``According to the transformations of SRT for a rest observer of
the stopwatch A the related with B momentarily co-moving inertial
chronometer has to, because of its relative inertial motion, measure
more slowly than the inertial stopwatch A. The circulating stopwatch
B measures for the rest observer of the stopwatch A the same rate
with its momentarily co-moving inertial chronometer. Consequently,
for the rest observer of the stopwatch A the circulating stopwatch
B has to measure \emph{at each moment} more slowly than the inertial
stopwatch A. Because the stopwatches are simultaneously activated
and simultaneously deactivated at their meetings, the result by the
rest observer of the stopwatch A is that after deactivation the stopwatch
B has to display less time elapsed than the stopwatch A. So for any
observer $t(B)<t(A)$.''

What does the sentence {}``\emph{for the rest observer of the stopwatch
A the circulating stopwatch B has to measure at each moment more slowly
than the inertial stopwatch A}'' mean? In SRT, one \textbf{cannot}
use the words {}``\emph{at each moment}'' with respect to different
co-moving inertial chronometers. It is well known that simultaneity
\textbf{is not} defined in SRT with respect to different co-moving
inertial chronometers \cite{key-2,key-3}. Clearly, the author of
\cite{key-1} misunderstands the meaning of the relativity of time
in SRT. The foundations of SRT imply that the rate of time measured
by an inertial moving observer is less than the rate of time measured
by an observer at rest. But the word \emph{rate} refers to \textbf{finite
intervals of times}, not to single moments! To understand this issue,
following \cite{key-2} let us consider the line-element in a flat
Lorentz-Minkowsi space-time 

\begin{equation}
ds^{2}=c^{2}dt^{2}-dx^{2}-dy^{2}-dz^{2},\label{eq: metrica lorenziana}\end{equation}

where $t,\mbox{ }x,\mbox{ }y,\mbox{ }z$ represents the 4-coordinates
of an observer $O$ at rest. Let us call $t',\mbox{ }x',\mbox{ }y',\mbox{ }z'$
the 4-coordinates of an inertial observer $O'$ moving with a constant
velocity $v$ with respect $O$. The invariance of the interval (\ref{eq: metrica lorenziana})
for inertial frames implies:

\begin{equation}
ds^{2}=c^{2}dt^{2}-dx^{2}-dy^{2}-dz^{2}=c^{2}dt'^{2},\label{eq: lorenziana 2}\end{equation}

as $dx'=dy'=dz'=0$ for the observer $O'$. Then

\begin{equation}
dt'=dt\sqrt{1-\frac{dx^{2}+dy^{2}+dz^{2}}{c^{2}dt^{2}}},\label{eq: trasformazione tempi}\end{equation}

and, as it is $v^{2}=\frac{dx^{2}+dy^{2}+dz^{2}}{dt^{2}}$, one gets

\begin{equation}
dt'=dt\sqrt{1-\frac{v^{2}}{c^{2}}},\label{eq: trasformazione 2}\end{equation}

that, integrated gives

\begin{equation}
t'_{2}-t'_{1}=\int_{t_{1}}^{t_{2}}dt\sqrt{1-\frac{v^{2}}{c^{2}}},\label{eq: trasformazione 3}\end{equation}

i.e. \begin{equation}
\triangle t'=\sqrt{1-\frac{v^{2}}{c^{2}}}\triangle t.\label{eq: trasformazione 4}\end{equation}

Clearly, Eqs. (\ref{eq: trasformazione 3}) and (\ref{eq: trasformazione 4})
\emph{cannot} be applied to two observers in a circular motion because
the line-element (\ref{eq: metrica lorenziana}) is invariant \textbf{only}
for inertial observers \cite{key-2,key-3}! In fact, even if it is
correct that in both of two arbitrary points 1 and 2 of a circular
motion one can associate two different momentarily co-moving inertial
observers, one \textbf{cannot} go from 1 to 2 through an inertial
motion because the motion is circular! Therefore, the transformations
of SRT (\ref{eq: trasformazione 3}) and (\ref{eq: trasformazione 4})
which take into account \textbf{finite} intervals of times\textbf{
cannot} be used by the author of \cite{key-1} for his wrong claims. 

In Section 2 of \cite{key-1} the author merely invert A with B in
his wrong demonstration by obtaining that, verbatim, {}``for any
observer $t(A)<t(B)$, what is false if for any observer $t(B)<t(A)$''
and he concludes that {}``The relative time dilatation of special
relativity theory has been mathematically disproved.''

Therefore, from the above analysis it is clear that Boltcho is totally
wrong and he demonstrated nothing.

On the other hand, it is well known that the experiment proposed by
Boltcho has been realized various times \cite{key-2} and the experimental
results have shown that it is the stopwatch B which measures more
slowly than the inertial stopwatch A. The inverse reasoning, in which
the role of the two stopwatch are inverted, is not correct because
the stopwatch B does not realize a straight uniform motion, i.e. it
is not an inertial observer \cite{key-2}. 

It is also important to emphasize that the wrong claims by Boltcho
on the supposed incorrectness of SRT in vacuum have nothing to do
with Santilli's criticisms on the necessity to modify SRT within Classical
Interior Dynamical Systems \cite{key-4}. SRT in vacuum is, perhaps,
the scientific theory which obtained the greatest number of experimental
tests in all the history of the human sciences. Nobody can tell that
it is wrong if they are confused by its foundations.

\subsubsection*{Acknowledgements}

I thank Arthur Boltcho for discussions on SRT. I also thank the Editor
of the Hadronic Journal, Prof. Jeremy Dunning-Davies, for considering
this work.

\end{document}